\documentclass[a4paper,fleqn,usenatbib]{mnras}

\usepackage{newtxtext,newtxmath}

\usepackage[T1]{fontenc}
\usepackage{ae,aecompl}

\usepackage{graphicx}	
\usepackage{subfigure}
\usepackage{amsmath}	
\usepackage{pdflscape}
\usepackage{float} 
\usepackage{here}
\usepackage{calc}

\title[Diameters of interferometric calibrators]{Improving the diameters of interferometric calibrators with MATISSE }

\author[S. Robbe-Dubois et al.]{
S. Robbe-Dubois$^{1}$\thanks{E-mail: sylvie.robbe-dubois@univ-cotedazur.fr},
P. Cruzal\`{e}bes$^{1}$,
Ph. Berio$^{1}$,
A. Meilland$^{1}$,
R.-G. Petrov$^{1}$,
\newauthor 
F. Allouche$^{1}$,
D. Salabert$^{1}$,
C. Paladini$^{2}$,
A. Matter$^{1}$,
F. Millour$^{1}$,
S. Lagarde$^{1}$,
B. Lopez$^{1}$,
\newauthor 
L. Burtscher$^{3}$,
W. Jaffe$^{3}$,
J. Hron $^{4}$,
I. Percheron$^{5}$, 
R. van Boekel$^{6}$,
G. Weigelt$^{7}$
and Ph. Stee$^{1}$
\\
$^{1}$Laboratoire Lagrange, Universit\'{e} C\^{o}te d'Azur, Observatoire de la C\^{o}te d'Azur, CNRS, Boulevard de l'Observatoire, CS 34229, F-06304 Nice Cedex 4, France\\
$^{2}$European Southern Observatory, Alonso de Cordova 3107, Vitacura, Santiago, Chile\\
$^{3}$Leiden Observatory, Leiden University, Niels Bohrweg 2, NL-2333 CA Leiden, the Netherlands\\
$^{4}$Department of Astrophysics, University of Vienna, T\"{u}rkenschanzstra\ss e 17, A-1180 Vienna, Austria\\
$^{5}$European Southern Observatory Headquarters, Karl-Schwarzschild-Stra\ss e 2, D-85748 Garching bei M\"{u}nchen, Germany\\
$^{6}$Max Planck Institut f\"{u}r Astronomie, K\"{o}nigstuhl 17, D-69117 Heidelberg, Germany\\
$^{7}$Max-Planck-Institut  f\"{u}r  Radioastronomie,  Auf  dem  H\"{u}gel  69,  D-53121 Bonn, Germany
}

\date{Accepted 2021 November 6. Received 2021 October 27; in original form 2021 September 23}

\pubyear{2022}

\begin{document}
\label{firstpage}
\pagerange{\pageref{firstpage}--\pageref{lastpage}}
\maketitle

\begin{abstract}
A good knowledge of the angular diameters of stars used to calibrate the observables in stellar interferometry is fundamental. As the available precision for giant stars is worse than the required per cent level, we aim to improve the knowledge of many diameters using MATISSE (Multiple AperTure mid-Infrared SpectroScopic Experiment) data in its different instrumental configurations. Using the squared visibility MATISSE observable, we compute the angular diameter value, which ensures the best fit curves, assuming an intensity distribution of a uniform disc. We take into account that the transfer function varies over the wavelength and is different from one instrumental configuration to another. The uncertainties on the diameters are estimated using the residual bootstrap method. Using the low spectral resolution mode in the $L$ band, we observed a set of 35 potential calibrators selected in the Mid-infrared stellar Diameter and Flux compilation Catalog with diameters ranging from about 1 to 3\,mas. We reach a precision on the diameter estimates in the range 0.6 per cent to 4.1 per cent. The study of the stability of the transfer function in visibility over two nights makes us confident in our results. In addition, we identify one star, 75~Vir initially present in the calibrator lists, for which our method does not converge, and prove to be a binary star. This leads us to the conclusion that our method is actually necessary to improve the quality of the astrophysical results obtained with MATISSE, and that it can be used as a useful tool for 'bad calibrator' detection.
\end{abstract}




\section{Introduction}

MATISSE (Multiple AperTure mid-Infrared SpectroScopic Experiment), a second generation instrument of the Very Large Telescope Interferometer (VLTI), is a combined imager and spectrograph for interferometry working in the mid-infrared 3--5\,$\mu$m ($L$ and $M$bands) and 8--13\,$\mu$m ($N$ band) spectral windows \citep[][]{Lagarde2012, Lopez2021} . The angular resolution in $L$ is about 3 mas and the spectral resolution ranges between 30 and 5000. The instrumental configuration is detailed in \citet[][]{Allouche2016}.

MATISSE builds on the experience gained with the VLTI's first generation instruments. It employs multi-axial beam combination with up to four Unit Telescopes (UTs) or Auxiliary Telescopes (ATs) of the VLTI while measuring the spectral distribution of visibilities, correlated flux, closure and differential phases, allowing aperture-synthesis imaging at different spectral resolutions. It will significantly contribute to several fundamental research topics in astrophysics \citep[][]{Wolf2008,Lopez2014,Matter2016,Lopez2018}, focusing for instance on the inner region of discs around young stars, where planets form and evolve, the surface structure and mass loss of stars at different evolutionary stages, and the environment of black holes in active galactic nuclei. 

After more than one decade of concept study, development, and manufacturing, MATISSE was fully integrated and tested at the Observatoire de la C\^{o}te d'Azur \citep[][]{Lopez2012,Robbe2014,Matter2016,Matter2016b}. In 2017~October, the instrument was shipped to Paranal, Chile, the ESO site of the VLTI. The alignment, integration and verification phase lasted from 2017~November to 2018~February, at the end of which first observations on the sky were performed to test the operations with the VLTI and to obtain the first stellar light \citep[][]{Robbe2018}. Numerous commissioning runs aiming at assessing the performance for the various instrument modes, on the different telescopes (UTs and ATs), and in an evolving VLTI infrastructure [GRA4MAT fringe tracker, Woillez et al. (in preparation)], have followed until June 2021. The performance on the sky are listed in \citet[][]{Petrov2020}.

To provide a reliable scientific exploitation of the MATISSE data, reference targets with well-known angular diameters and accurate absolute visibilities are essential to calibrate the interferometric observables. For this goal, \citet{Cruzalebes2019} built an all-sky catalogue called the Mid-infrared stellar Diameter and Flux compilation Catalogue (MDFC, Cat. II/361), compiling data helpful for mid-infrared interferometry with about 453\,000 stellar sources contained in the JMMC Stellar Diameters Catalogue \citep[JSDC, Cat. II/346,][]{Bourges2017}. The angular diameters reported in the JSDC derive from an improvement of the surface brightness method \citep[][]{Wesselink69,Barnes76}. \citet[][]{Chelli2016} introduced a reddening-free self-calibrated observable (the differential surface brightness, DSB), independent of the distance, and measured via photometry and interferometry. The polynomial fit of the DSB as a function of the spectral type number for a data base of about 600 stars with known diameters and photometries allowed them to compute the angular diameter of the stars of the JSDC catalogue with known spectral types and \textit{VJHKs} magnitudes.

Out of these sources,'IR-excess-free' calibrators were identified as especially suitable for the mid-infrared spectral domain \citep[see][for a description of the IR excess-free criterion]{Cruzalebes2019}. For observations with the 1.8m Auxiliary Telescopes of the VLTI, that list includes: 1\,621 bright calibrators are proposed for the $L$ band only, 44 bright and medium-bright calibrators for the $N$ band only, while 375 'hybrid' bright, medium-bright, and faint calibrators suitable for both spectral bands. For observations with the 8-m Unit Telescopes, the list includes 259 bright calibrators suitable for both spectral bands. 

The angular diameters of these sources (mainly K-giants) are not given in the JSDC with a better precision than 8 per cent to 10 per cent. The less accurate the calibrator diameter, the higher the uncertainty on the calibrated data. That can lead to a loss of pertinence on the results or even astrophysical misinterpretation. It thus appears necessary to derive accurate angular diameters for those IR calibrators, and establish the most robust strategy to 'calibrate the calibrators'. In this article, we present a new approach to measure calibrator angular diameters from MATISSE data with a precision down to 1 per cent. We present the results obtained on MATISSE data acquired in the frame of a dedicated observing campaign in 2019 and 2020. 

The methodology to process the data to achieve this goal is presented in Section~\ref{sec:method}, and the results obtained for a first set of calibrators observed by ESO in 2019 and the beginning of 2020 in Section~\ref{sec:results}. Section~\ref{sec:belief} presents  examples of measurements that reinforce the reliability in these results, and Section~\ref{sec:inspection} shows how this method can be used as a good calibrator inspection.


\section{Methodology} \label{sec:method}

The method consists in deriving the stellar angular diameter from one MATISSE observable, the spectral distribution of the squared visibility $V^2(\lambda)$, by taking properly into account the shape of the transfer function (instrument + atmosphere) and by assuming that the calibrator intensity distribution is well represented by a uniform disc (UD). The proposed methodology is based on the double assumption that both the $V^2(\lambda)$ contribution of the calibrator and the transfer function are sufficiently well-described by simple parametric models. The first assumption is trivial, as we consider UDs. The second assumption is not so trivial as the transfer function includes the Earth's atmosphere variations, in addition to the MATISSE and the VLTI instrumental responses. It is the reason why we developed a method based on a 'self-calibration' of the instrument using the available different instrumental configurations.

\subsection{Instrumental configuration}\label{sec:inst}

The software of MATISSE provides simultaneous measurements of the squared visibilities $V^2(B, \lambda)$ associated with the six interferometric baselines $B$ given by each VLTI configuration \citep[][]{Millour2016}. The baselines can be successively rearranged into four different configurations thanks to an internal instrumental module, the Beam Commuting Device (BCD), allowing the commutation of the beams by pairs \citep[][]{petrov2007}. 

When the BCD is in the 'OUT/OUT' mode, the four input beams (called IP1, IP3, IP5, and IP7) coming from the VLTI telescopes go through the instrument up to the detector with no beam commutation. When the BCD is in the 'IN/IN' mode, IP1 and IP3 are commuted, such as IP5 and IP7. This generates a change of the fringe peak positions in the Fourier plane. Intermediates modes are also available:~'IN/OUT' allowing the commutation between IP1 and IP3 but not between IP5 and IP7, and 'OUT/IN' allowing the commutation between IP5 and IP7 but not between IP1 and IP3. Originally, the function of these commutations is to minimize the instrumental effects on the phase measurements (closure phase and differential phase). In the $L$ band, in which fringes and photometric images must be recorded simultaneously to ensure accurate visibility measurements, the automatic sequence of observation is the following \citep[][]{Lopez2021}:
\begin{itemize}
    \item	two sky exposures of 30s recorded in the IN/IN and OUT/OUT BCD configurations,
    \item	four interferometric + photometric exposures of 60s for each of the BCD configurations (IN/IN, OUT/IN, IN/OUT and OUT/OUT) without telescope chopping (used to separate the stellar flux from the sky background),
    \item	2 x 4 interferometric + photometric exposures of 60s for each of the BCD configurations (IN/IN, OUT/IN, IN/OUT and OUT/OUT) with chopping.
\end{itemize}

In order to discriminate in the $V^2(B,\lambda)$ measurements the contribution of the calibrator considered as a UD, from the contribution of the transfer function, we take advantage of these different available configurations which provide four times more spectral data sets of $V^2(B,\lambda)$. 

The observations are made in $L$ at low spectral resolution ($R\approx30)$, in the selected spectral range 3.25--3.75\,$\mu$m, where the response in $V^2(B, \lambda)$ is, by experience, the less noisy. In this range at this resolution, each data set is composed of 21 spectral channels. An automatic spectral binning on 5 pixels (corresponding to the width of a spectral channel) is done by the data reduction pipeline.

\subsection{Diameter estimation}

\subsubsection{Customized observable} \label{sec:custom}
The question addressed in this section is to determine the most appropriate scalar function to run the minimization procedure, the variables of the function being the stellar angular diameter and the parameters describing model of the transfer function. It turned out that we do not directly fit the raw observable $V^2(B,\lambda)$.

First, to limit the number of parameters describing the transfer function model, we compute the referenced squared visibility such that:
\begin{equation}
    V^2_{\rm ref}(B, \lambda)=\frac{V^2(B, \lambda)}{V^2(B, \lambda_{\rm ref})},
	\label{eq:eq1}
\end{equation}
where $\lambda_{\rm ref}$ is the referenced wavelength taken at the middle of the selected spectral range in the flux continuum.

The associated transfer function $T_{\rm ref}(B, \lambda)$ is given by:
\begin{equation}
    T_{\rm ref}(B,\lambda)=V^2_{\rm ref}(B, \lambda)\frac{V^2_{\rm UD}(B, \lambda_{\rm ref})}{V^2_{\rm UD}(B, \lambda)},
	\label{eq:eq2}
\end{equation}
with $V_{\rm UD}(B, \lambda)$ the visibility for the UD model:
\begin{equation}
    V_{\rm UD}(B, \lambda)=\left|\frac{2 J_{1}(z)}{z}\right|,
	\label{eq:eq3}
\end{equation}
where $z = \pi \phi B / \lambda$ with $\phi$ is the angular diameter and $J_{1}$ the Bessel function of first order. 

To describe the wavelength dependence of $T_{\rm ref}$, we arbitrarily adopt a simple empirical polynomial law. Noting that the initial minimization procedure (described in the next section) does not converge for all the stars, we rather use $V^2(B, \lambda)$ divided by $V^2(B_{\rm short}, \lambda)$ measured at the shortest baseline:
\begin{equation}
    \tilde{V}^2_{\rm ref}(B, \lambda)=\frac{V^2_{\rm ref}(B, \lambda)}{V^2_{\rm ref}(B_{\rm short}, \lambda)},
	\label{eq:eq4}
\end{equation}
with the associated transfer function $\tilde{T}_{\rm ref}(B, \lambda)$ given by:
\begin{equation}
    \tilde{T}_{\rm ref}(B,\lambda)=\tilde{V}^2_{\rm ref}(B, \lambda)\frac{\tilde{V}^2_{\rm UD}(B, \lambda_{\rm ref})}{\tilde{V}^2_{\rm UD}(B, \lambda)}.
	\label{eq:eq5}
\end{equation}

\begin{figure*}
\centering
\includegraphics[width=0.6\textwidth]{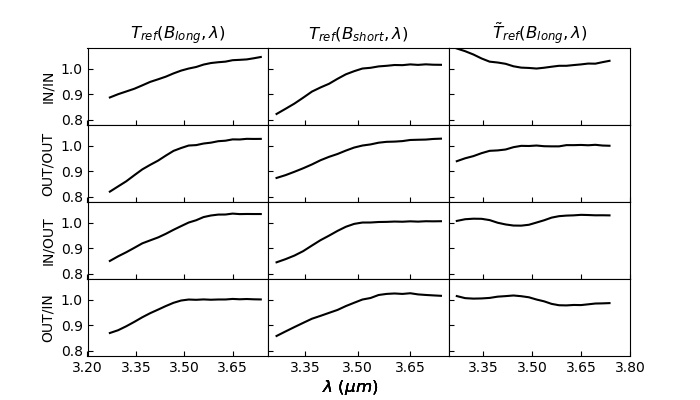}
    \caption{Observation of the star 35 Vir for the 4 BCD modes. Left-hand panels: transfer function $T_{\rm ref}(B_{\rm long},\lambda)$ obtained with a long baseline (132.4\,m). Middle panels: $T_{\rm ref}(B_{\rm short},\lambda)$ obtained with the shortest baseline (58.2\,m). Right-hand panels: ratio between both transfer functions $\tilde{T}_{\rm ref}(B_{\rm long},\lambda)~=~T_{\rm ref}(B_{\rm long},\lambda)$~/~$T_{\rm ref}(B_{\rm short},\lambda)$. }
    \label{fig:VBs}
\end{figure*}

Since the polynomial law better fits $\tilde{T}_{\rm ref}(B,\lambda)$ than $T_{\rm ref}(B,\lambda)$, the minimization process succeeds now for any star. To illustrate that, we show in Fig.~\ref{fig:VBs} the plot of $T_{\rm ref}(B_{\rm long},\lambda)$ obtained with a long baseline (132.4\,m) on the left, $T_{\rm ref}(B_{\rm short},\lambda)$ obtained at the shortest baseline (58.2\,m) on the middle, and the ratio $\tilde{T}_{\rm ref}(B_{\rm long},\lambda)~=~T_{\rm ref}(B_{\rm long},\lambda)$~/~$T_{\rm ref}(B_{\rm short},\lambda)$ on the right, derived from the observation of the star 35 Vir for the 4 BCD modes. Since the right and middle plots show curves with a linear increase as a function of $\lambda$ before 3.5\,$\mu$m or more, and a plateau after, they are obviously not described by a polynomial law for any baselines and BCD. The right plots show flatter curves closer to polynomial laws. 

$\tilde{V}^2_{\rm ref}(B, \lambda)$ is our 'customized observable' in the procedure described hereafter. To simplify the notations, we forget the tilde symbol in the following sections.

\subsubsection{Method for diameter estimation}\label{sec:meth}
In this section we describe the method used to estimate the stellar diameter $\phi$.
The method uses the residuals $\rho(B,\lambda)$ of the difference between the $V^2_{\rm ref}(B, \lambda)$ customized observable and its parametric model $V^2_{\rm model}(B,\lambda)$, defined as:
\begin{equation}
	\rho(B,\lambda) =\frac{V^2_{\rm ref}(B,\lambda)- V^2_{\rm model}(B,\lambda)}{\sigma_{\rm ref}(B,\lambda)}
	\label{eq:eq6}
\end{equation}
where $\sigma_{\rm ref}(B,\lambda)$ is the uncertainty of $V^2_{\rm ref}(B,\lambda)$ calculated from the errors on the measured squared visibility $V^2(B,\lambda)$ provided by the software of MATISSE. The squared visibility $V^2_{\rm model}(B, \lambda)$ is given by:
\begin{equation}
	  V^2_{\rm model}(B, \lambda) = T(B, \lambda)\frac{ V^2_{\rm UD}(B, \lambda)~/~ V^2_{\rm UD}(B, \lambda_{\rm ref})}{ V^2_{\rm UD}(B_{\rm short}, \lambda)~/~V^2_{\rm UD}(B_{\rm short}, \lambda_{\rm ref})}.
	\label{eq:eq7}
\end{equation}

where $T(B, \lambda)$ is the associated transfer function described by a quadratic function with baseline-dependent coefficients ($b$, $c$) so that:
\begin{equation}
	T(B,\lambda) = 1 + b(B) (\lambda-\lambda_{\rm ref}) + c(B) (\lambda-\lambda_{\rm ref})^2.
	\label{eq:eq8}
\end{equation}
Experience shows that the quadratic law better fits to the data, with lower $\chi^2$ values, than the linear law. Higher order polynomials do not provide lower $\chi^2$ values, while the computing time stays somewhat longer as it depends on the number of unknowns.

The process is the following:
\begin{enumerate}
\item We remove the extreme outliers (with the conservative threshold k~=~3.5 chosen to select the data) from the initial data set using the double MAD (Mean Absolute Deviation) approach \citep{Rosen} in which the residuals $\rho(B,\lambda)$ are computed with the stellar diameter reported in the JSDC catalogue and the transfer function $T(B,\lambda)$ equal to unity. 
\item We make a first estimation of the parameters of the transfer function given by Eq.~(\ref{eq:eq8}) with the same JSDC stellar diameter value, and with the data set corrected from the outliers. This first estimation, based on $\chi^2$ [i.e. $\rho(B,\lambda)$] minimization, allows us to run a second extreme outlier removal from the complete initial data set using again the double MAD approach on the resulted residuals. 
\item Once this removal is performed, we run a second $\chi^2$ minimization, which allows us to determine the sets of parameters $(\phi, b , c)$ producing the best fit of $V^2_{\rm ref}(B, \lambda)$ for each baseline. This is not fully satisfactory as the stellar diameter should, of course, not depends on the baseline. 
\item We perform a third and last $\chi^2$ minimization in order to find the best diameter value fitting all the baselines simultaneously, the sets of transfer function parameters $(b , c)$ being fixed to their previous values. 
\item At the end of this process, we perform a third and final extreme outlier removal before running the process of the diameter error estimation as described in Section~\ref{sec:acc}. At this point, the final data set results from the two extreme outlier removals applied in steps (ii) and (v).
\end{enumerate}

\subsubsection{Example of the angular diameter estimation: 35~Vir}

\begin{figure*}
\centering
\includegraphics[width=\textwidth]{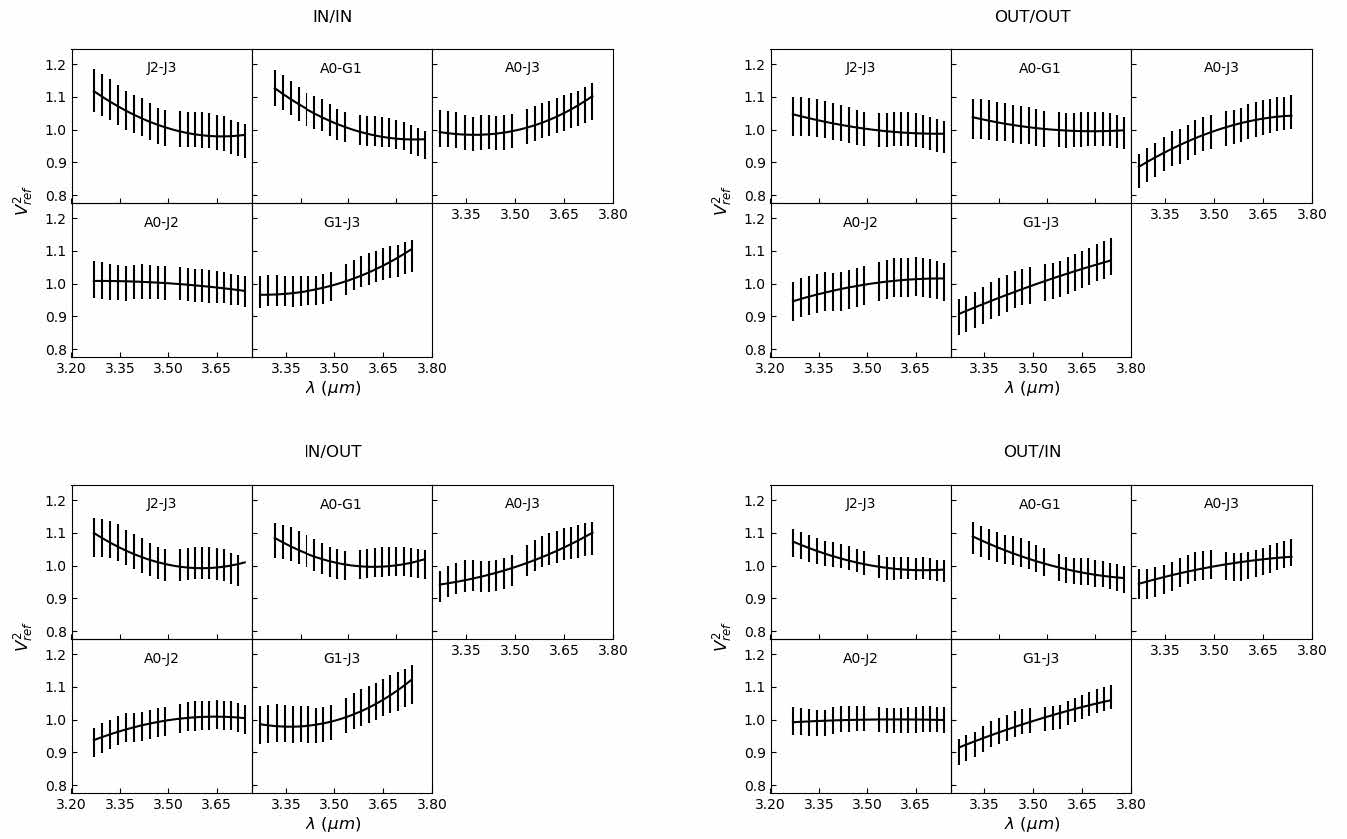}
    \caption{Customized observable $V^2_{\rm ref}(B, \lambda)$ obtained with the star 35~Vir ($\phi_{\rm JSDC}=2.54\pm0.22$\,mas) in the 4 BCD modes for each VLTI baseline. The solid black line is the model observable $V^2_{\rm model}(B, \lambda)$. The shortest baseline G1-J2 for which $V^2_{\rm ref}(B, \lambda)=1$ and is not represented here.}
    \label{fig:V2}
\end{figure*}

As an example, we consider the M0/1III red giant star 35~Vir.  We show in Fig.~\ref{fig:V2} the customized observable $V^2_{\rm ref}(B, \lambda)$  fitted with the model $V^2_{\rm model}(B, \lambda)$. The notations A0, G1, J2, J3 indicate the positions of the AT telescopes in the used VLTI configuration. 
The baseline lengths are equal to: 58.2\,m for G1--J2 ($B_{\rm short}$); 90.5\,m for A0--G1; 104.0\,m for J2--J3;  129.3\,m for A0--J2; and 132.4\,m for both A0--J3 and G1-J3. The angular diameter value derived with our method is 2.51\,mas, against the value of $2.54\pm0.22$\,mas reported in
the JSDC catalogue.

\begin{figure*}
\centering
	\includegraphics[width=0.7\textwidth]{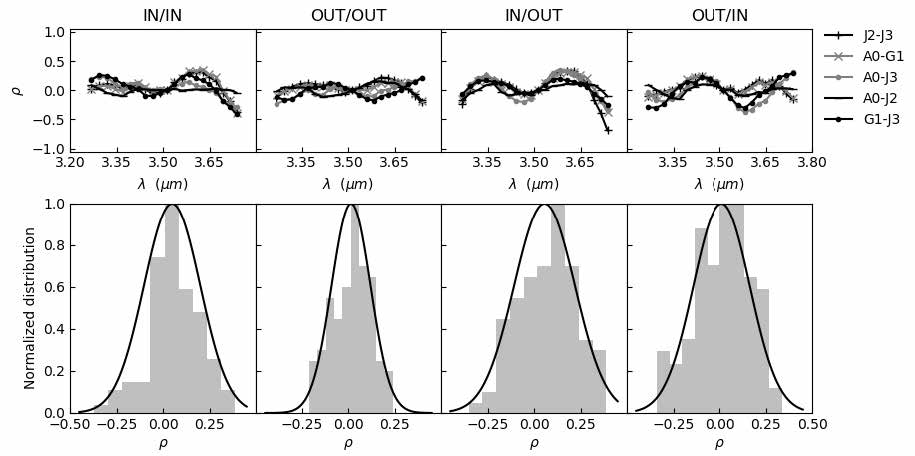}
    \caption{Residuals $\rho(B, \lambda)$ as a function of the wavelength and  distributions for each BCD mode obtained with 35~Vir. The black solid curve is the normal law shown for comparison. }
    \label{fig:res}
\end{figure*}

To emphasize the good quality of the fit, we also show in Fig.~\ref{fig:res} the spectral distribution of the residuals $\rho(B,\lambda)$, and their statistical distributions, which are close to normal distributions. 

\begin{figure*}
\centering
\includegraphics[width=\textwidth]{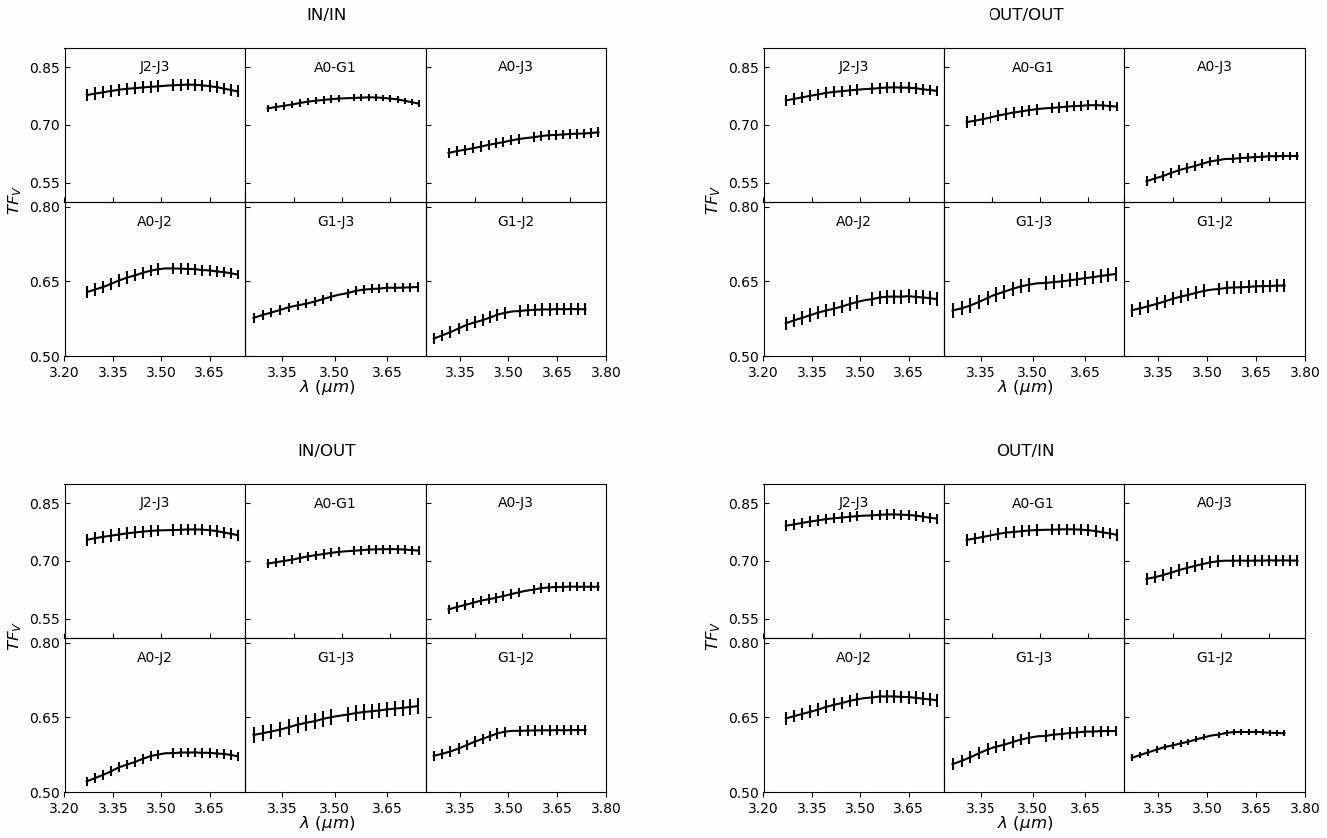}
    \caption{Transfer function in visibility deduced from the ratio between $V(B, \lambda)$ and $V_{\rm UD}(B, \lambda)$ obtained with the star 35~Vir in the 4 BCD modes for each baseline. }
    \label{fig:FT}
\end{figure*}

In Fig.~\ref{fig:FT}, we show the transfer function $\rm TF_V$ in visibility deduced from the ratio between $V(B, \lambda)$ and the UD model $V_{\rm UD}(B, \lambda)$ [Eq.~(\ref{eq:eq3})]. This allows to identify the variations between the BCD modes, mainly due to the spatial coding of the fringe peaks in the Fourier plane (see Section~\ref{sec:inst}), and needs to be taken into account in the data processing with our method.


\subsection{Precision estimation with the residual-bootstrap method} \label{sec:acc}

\subsubsection{Methodology}
We derive robust estimates of the statistical uncertainties of the model parameters from the confidence limits given by the bootstrapping procedure with residual resampling \citep[described in Appendix C of][]{Cruz2010}. The process is the following:
\begin{enumerate}
	\item We centre the residuals by subtracting the mean of the residual terms and resample the centred residuals by drawing randomly with replacement, following a uniform distribution, so that a new residual value $\rho_{\rm new}(B, \lambda)$ is obtained for each measurement.
	\item We build a new synthetic data set defined as:
\begin{equation}
	V^2_{\rm synth}(B, \lambda)=V^2_{\rm model}(B, \lambda) + {\sigma_{\rm ref}(B,\lambda)}~\rho_{\rm new}(B, \lambda).
	\label{eq:eq9}
\end{equation}
	\item For one sample of the synthetic data set, we run the two $\chi^2$-minimizations [points (iii) and (iv) of Section~\ref{sec:meth}], resulting in transfer function parameter estimation and angular diameter determination. If $N$ is the number of derived synthetic data sets, we obtain a final set of $N$ values of ($\chi^2$, $\phi$).
	\item Finally, we search for extreme outliers in diameter, using again the double MAD method. It turns out that all the $N$ points ($\chi^2$, $\phi$) are kept for the estimation of the uncertainty on the stellar diameter, which is simply the half range calculated with the two extreme values. 
\end{enumerate}

\subsubsection{Example of the estimation the angular diameter precision: 35~Vir}
\begin{figure}
\centering
	\includegraphics[width=\columnwidth]{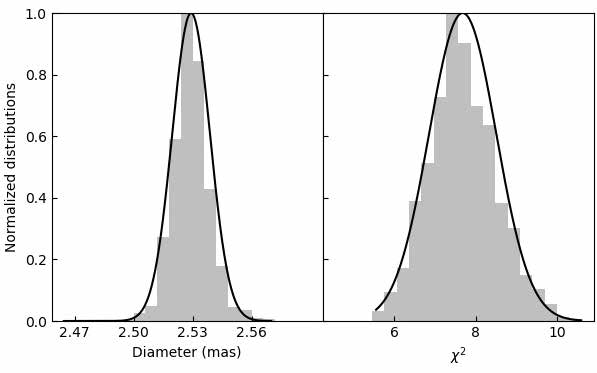}
    \caption{Angular diameter and $\chi^2$ distributions obtained by bootstrapping with 35~Vir. }
    \label{fig:boot}
\end{figure}

In Fig.~\ref{fig:boot}, we show the $\phi$ and $\chi^2$ statistical distributions obtained with the residual-bootstrap method. The large number of repeats ($N~=~1000$) provides a good statistics and reproducibility.
The resulting error on the angular diameter is 0.05\,mas, resulting in a relative error of 2 per cent to be compared with the relative error of 8.8 per cent provided by the initial values reported in the JSDC ($\phi_{\rm JSDC}=2.54\pm0.22$\,mas).

\section{Results}
\label{sec:results}
\subsection{Calibrator selection}

From the MDFC catalogue, we extracted a primary list of 'pure' calibrator candidates suitable for the $L$-band observable by MATISSE with the ATs. The sources must be:
\begin{itemize}
    \item observable from the ESO-Paranal observatory; 
    \item classified as potential calibrators in the JSDC (reliable diameter estimate, 'favourable' object type, no close binary);
    \item partially resolved with the longest baseline of the VLTI (130\,m), hence with an angular diameter estimate less than 3\,mas ensuring the calibration error caused by the uncertainty in the calibrator modelling to be minimized \citep{Cruzalebes2019};
    \item brighter than 10\,Jy in correlated flux (in L) at the longest baseline; 
    \item with a precision in flux better than 15 per cent computed with at least two photometric values (in L) reported in the catalogue; and
    \item unsuspected to show any excess, extent, or variability in the mid-infrared spatial domain. 
\end{itemize} 
We found 171 sources fulfilling these criteria. About 75 per cent of them are identified as cool giants with SIMBAD (G, K, and M spectral types). Three sources are MIDI calibrators \citep{Verhoelst2005} and five are spectro-photometric standards according to \citet{Cohen1999}. Around fifty sources of this list have an estimated flux higher than 20\,Jy in L. 

\begin{figure}
\centering
	\includegraphics[width=\columnwidth]{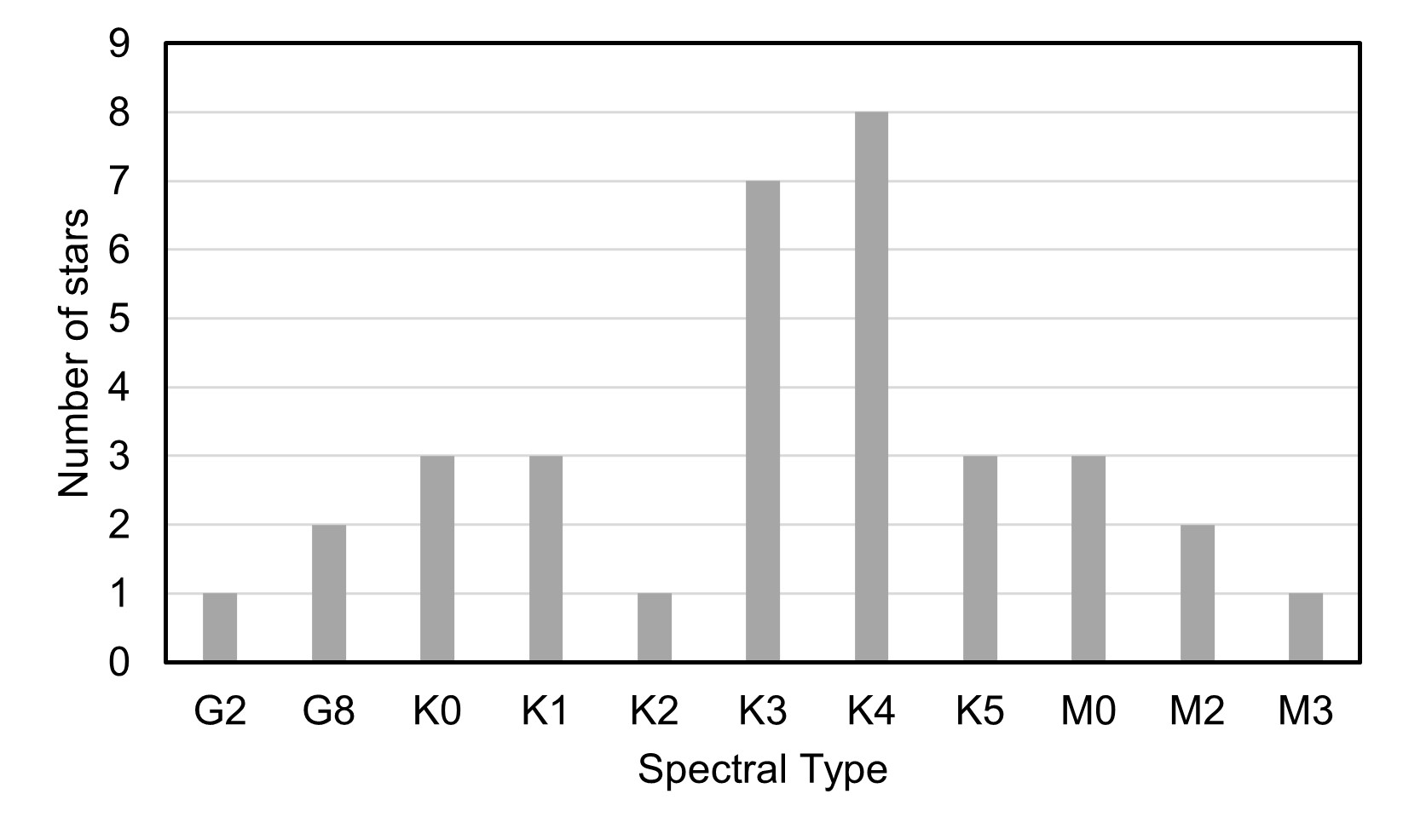}
    \caption{Distribution in spectral type of the observed stars. }
    \label{fig:sptype}
\end{figure}

\begin{figure*}
\centering
	\includegraphics[width=0.92\textwidth]{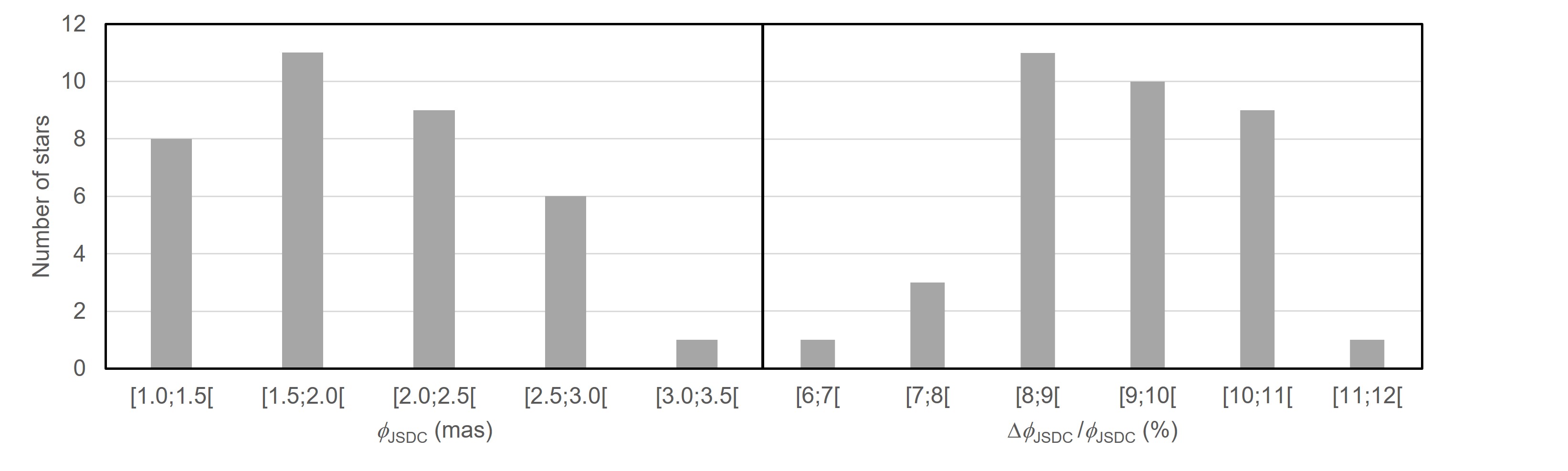}
    \caption{Distribution in angular diameter (left-hand panel) and relative uncertainty (right-hand panel) reported in the JSDC catalogue for the proposed stars.}
    \label{fig:jsdcdiam}
\end{figure*}

In Table~\ref{tab:calib}, we give the list of first targets that we observed within the framework of our project. We also included four very bright stars not identified as 'pure' potential IR calibrators, i.e. IRFLAG $\neq$ 0 \citep[see][]{Cruzalebes2019}. In Fig.~\ref{fig:sptype}, we show the distribution in spectral type of those stars. In Fig.~\ref{fig:jsdcdiam}, we show the distribution in angular diameter reported in the JSDC catalogue. The angular diameter values range from 1.05 to 3.10\,mas, with relative uncertainties ranging from 6.8 to 11.4 per cent, 9.2 per cent on average.

\subsection{Resulting angular diameters and accuracies}

As shown in Table~\ref{tab:result}, our MATISSE $L$-band observations were performed from May~2019 to February~2020 in low spectral resolution, using the large AT array (A0--G1--J2--J3) and its variation (A0--G1--J2--K0). One observation was carried out with the hybrid configuration A0--B2--D0--J3. Since the observations with ATs failed with $\iota$~Ant, we therefore use a UT observation taken in the ESO archive data base.

In Table~\ref{tab:result}, we show the angular diameters with their uncertainties derived from our method. Several stars were observed twice:~$\chi$~Eri and HD~189695 in the same night, $\mu$~Psc and HD~184996 in two different nights, thus allowing to test the reproducibility of the method. As shown in Section~\ref{sec:inspection}, the star 75~Vir cannot be described by a UD and is therefore excluded from the table.

\begin{figure*}
\centering
	\hspace{-0.5em}\includegraphics[width=0.91\textwidth]{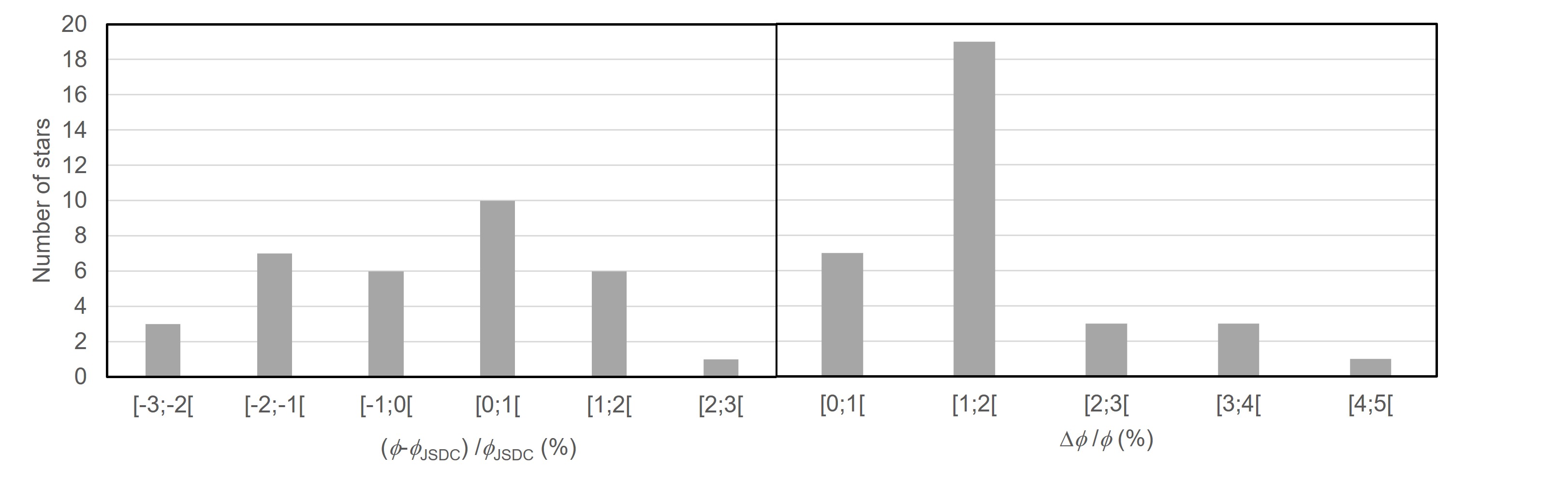}
    \caption{Left-hand panel: distribution of the relative difference in angular diameter of our final values with respect to the JSDC values. Right-hand panel: distribution in relative uncertainty of the diameters derived from our method.}
    \label{fig:findiam}
\end{figure*}

In Fig.~\ref{fig:findiam}, we show the relative difference in angular diameter of our calculated final values with respect to the reported JSDC values and the relative uncertainty on the final diameters. The mean of the absolute values of the relative difference is about 1 per cent, while the relative uncertainties range from 0.6 to 4.1 per cent, 1.6 per cent on average. 

For the two lowest precision values, the minimization did not converge for each baseline [step (iii) of Section~\ref{sec:meth}], resulting on a total number of analysed data points of 30 per cent and 40 per cent of the complete initial data set. We also noted that the number of outliers removed at steps (ii) and (v) of the procedure concerns most of the times less than 2 per cent the initial data set. In the worst case of the 8 per cent rejected data we encountered, the difference in the diameter estimation is 0.25 per cent when the outliers are not removed. 

\section{Confidence in the diameter knowledge} \label{sec:belief}
\subsection{Reproducibility}
The consecutive observations of the four following stars show quite reproducible results (less than 1 per cent):
\begin{itemize}
    \item $\mu$~Psc: $2.461\pm0.020$\,mas (2019 July 20) and $2.462\pm0.022$\,mas (2019 July 21)
    \item $\chi$~Eri: $2.055\pm0.016$\,mas and $2.067\pm0.015$\,mas (2019 July 20)
    \item HD~184996: $2.656\pm0.024$\,mas (2019 July 21) and $2.676\pm0.016$\,mas (2019 July 23)
    \item HD~189695: $2.041\pm0.059$\,mas and $2.023\pm0.069$\,mas (2019 May 19)
\end{itemize}

\subsection{Uniform versus limb-darkened disc model}
Our method of diameter estimation is based on the UD model. This section questions the relevance of this hypothesis by comparing the resulting angular diameters with a limb-darkened disc (LD) model. Using the linear law described by \citet{han}, we use the parametric expression given by:
\begin{equation}
	V_{\rm LD}(B,\lambda) =\left( \frac{6}{3-u_{\lambda}}\right) \left|  (1-u_{\lambda})~\frac{ J_{1}(z)}{z} + u_{\lambda}\sqrt{\frac{\pi}{2}}~\frac{J_{3/2}(z)}{z^{3/2}} \right|.
	\label{eq:eq10}
\end{equation}
where $u_{\lambda}$ is the limb darkening parameter (0$\leq u_{\lambda} \leq$1), and $J_{3/2}$ the Bessel function of order 3/2. In this study, we consider that $u_{\lambda}$ is constant and equal to $u$ over the wavelength range of the $L$ band.

\begin{figure*}
\centering
	\includegraphics[width=0.7\textwidth]{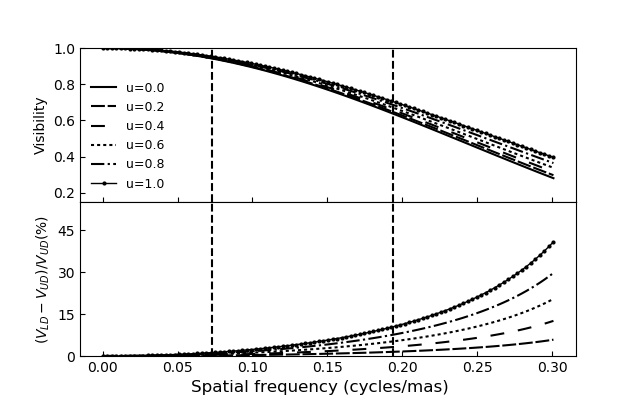}
    \caption{Top panel: visibility versus spatial frequency obtained for different values of $u$. Bottom panel: relative deviation from the UD visibility. The UD's angular diameter is 3\,mas. The dashed vertical lines indicate the extreme values for the spatial frequency range accessible in $L$ with the used AT configurations.}
    \label{fig:UDvsLD}
\end{figure*}

Fig.~\ref{fig:UDvsLD} shows the visibility expected from Eq.~(\ref{eq:eq10}) versus the spatial frequency obtained with different values of $u$, and the relative deviation from the UD $(V_{\rm LD}-V_{\rm UD})/V_{\rm UD}$ obtained with an angular diameter of 3\,mas. The dashed vertical lines indicate the extreme values of the frequency range accessible in $L$ with the used AT configurations. At the highest frequency of about 0.2\,cycles.mas$^{-1}$ (baseline of 132\,m), the LD-visibility exceeds the UD-visibility by 10 per cent for the same angular diameter.

We run our procedure described in Section~\ref{sec:meth} using now the linear LD-model. To study the effect of the limb-darkening on the diameter estimation, we vary the value of the $u$ coefficient of Eq.~(\ref{eq:eq10}) in its full range between 0 and 1. 

\begin{figure*}
\centering
	\includegraphics[width=0.7\textwidth]{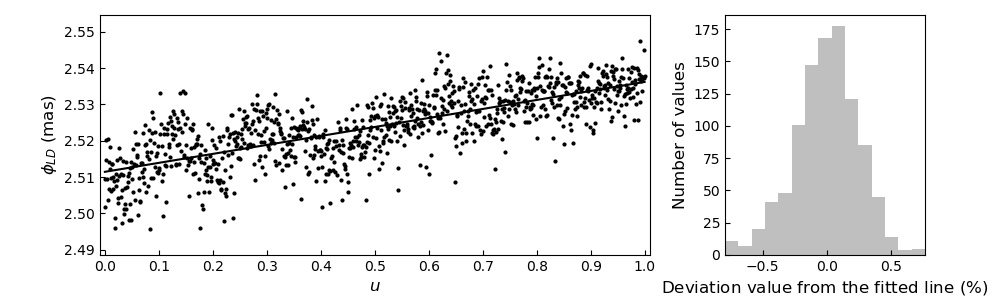}
    \caption{Left-hand panel: angular diameters $\phi_{LD}$ of 35~Vir using a linearly LD versus the darkening parameter $u$. Right-hand panel: distribution of the deviation between the angular diameter for a given $u$ and the value on the fitted straight line.}
    \label{fig:LD}
\end{figure*}

Figure~\ref{fig:LD} shows the resulted angular LD-diameters $\phi_{\rm LD}$ obtained for 35~Vir with our method. For this calibrator, the dispersed values due to the limited precision inherent in the minimization method are properly fitted by a straight line with a positive slope. Table~\ref{tab:ass} gives for each calibrator the relative deviation between the UD diameter $\phi_{\rm UD}$ (Table~\ref{tab:result}) and the fitted LD diameter $\phi_{0}$ at $u~=~0$, the ratio $\sigma$ between the root mean square error of the distribution and $\phi_{0}$, and the maximal deviation between the two fitted angular diameter extreme values $\phi_{1}$ at $u~=~1$, and $\phi_{0}$. 

Except for HD~94014 and HD~184996, the values of the deviation are lower than the precision of our method considering the UD (Table~\ref{tab:ass}). We have investigated whether these two calibrators present a limb-darkening profile or not. We run our programme using the LD-model, this time with two unknowns: the angular diameter $\phi_{\rm LD}$ and the parameter $u$. The minimization process ends up with a deviation from $\phi_{\rm UD}$ of 0.006 per cent and 0.3 per cent respectively, and $u$ equal to 0.01 and 0.02. This confirms that these two stars are well described by a UD model. We conclude that our hypothesis is well adapted to our method, which is in fact limited by the precision in the calculations.

\subsection{Calibrated visibilities}
To illustrate the reliability of our method, we applied it to the estimation of the transfer function in visibility ($\rm {TF}_V$) of one full night of MATISSE observation.
For that, we selected one night (2018 December 09) from the MATISSE imaging commissioning run with fairly good seeing (between 0.5 and 1 arcsec) and coherence time (between 5 and 10\,ms). Two science targets (the giant carbon star R~Scl and the B[e] star  F~ CMa) and three calibrators (7~Cet, $\pi$~Eri, $\varepsilon$~Lep) were observed during that night.

Table~\ref{tab:calibTF} gives the diameters taken from the JSDC data base and those estimated using our method. Their values are compatible, but the uncertainties are about 10 times smaller using our method.

\begin{figure*}
\centering
	\includegraphics[width=0.4\textwidth]{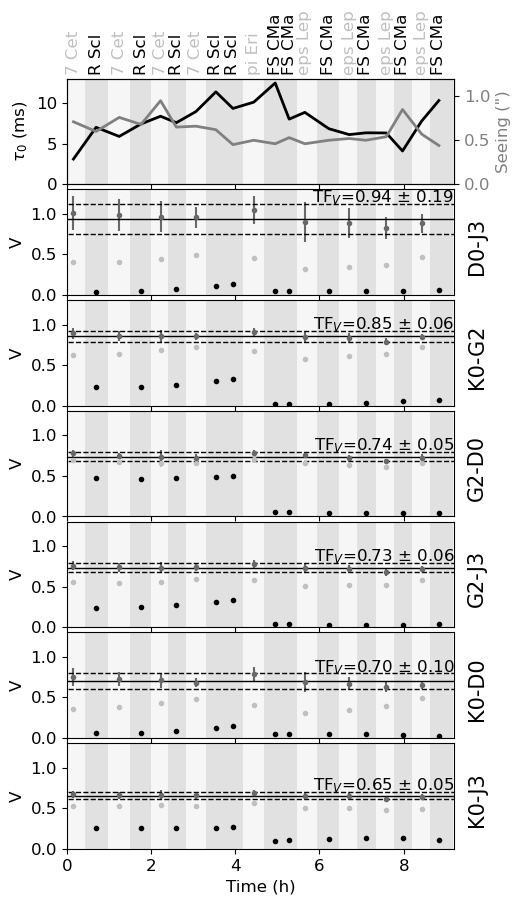}
	\includegraphics[width=0.4\textwidth]{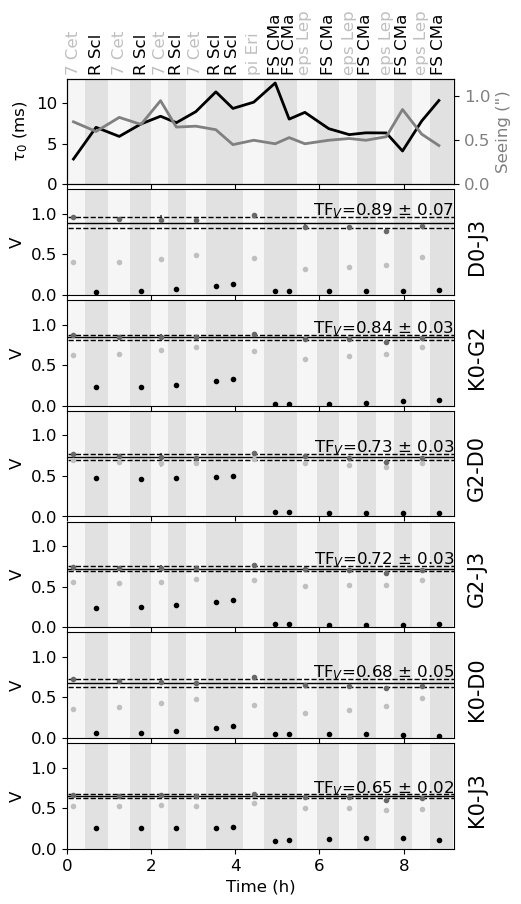}	
    \caption{Mean $L$ band visibility $\rm V$ (averaged between 3.1 and 3.8 $\mu$m) plotted as a function of time for a full night (2018 December 09), and the six VLTI baselines for the BCD OUT/OUT. The two science objects, FS CMa and R Scl, are shown in black, and the calibrators are in light gray. The estimated transfer function $\rm TF_V$ in visibility (i.e. corrected from the partial resolution of the calibrators) is shown in dark gray. In the left-hand panel, the JSDC diameters values are used, whereas more realistic values obtained with our new estimations are shown in the right-hand panel. The horizontal line represents the mean value of visibility during the night (solid line) and its standard deviation (dashed line).
    }
    \label{fig:Vis_anthony}
\end{figure*}

In Fig~\ref{fig:Vis_anthony}, we plot the mean $L$ band visibility and transfer function in visibility $\rm TF_V$ as a function of time, for the BCD OUT/OUT. $\rm TF_V$  (=~$V/V_{\rm cal}$ with $V_{\rm cal}$ the calibrator visibility) is computed using both estimation of the diameters. Using our method, we obtain a more stable visibility with a nightly standard deviation at least twice smaller on all baselines. This is not only due to the smaller errors on our estimation of the diameters. On the JSDC values, the large standard deviation is mainly due to the spread between values taken from the three different calibrators. This confirms that our method allows the fine-tuning of the knowledge on the diameters of the calibrators used to calibrate the visibilities.

\begin{table}
\renewcommand\thetable{4}
\caption{Diameters for the three calibrators observed during the MATISSE imaging commissioning night of 2018 December 9th, taken from the JSDC database and estimated with our method.}       
\label{tab:calibTF}                              
\centering\begin{tabular}{ccc}        
\hline \hline  
NAME & \multicolumn{2}{c}{ Estimated diameter (mas)}\\
 & JSDC & This paper\\
\hline\
7 Cet & 5.50$\pm$0.55 & 5.37$\pm$0.05\\
$\pi$  Eri & 5.28$\pm$0.47&5.09$\pm$0.06\\
$\varepsilon$ Lep &5.87$\pm$0.56 & 5.71$\pm$0.07\\
\hline
\end{tabular}
\end{table}

\subsection{Mean transfer function stability}

We studied then the stability of the mean transfer function $\rm TF_V$ (=~$V/V_{\rm UD}$) in visibility over one night of observation of the calibrators. The mean is taken over the wavelengths, which averaged between 3.1 and 3.8\,$\mu$m as in the previous Section. 

\begin{figure*}
\centering
	\includegraphics[width=0.8\textwidth]{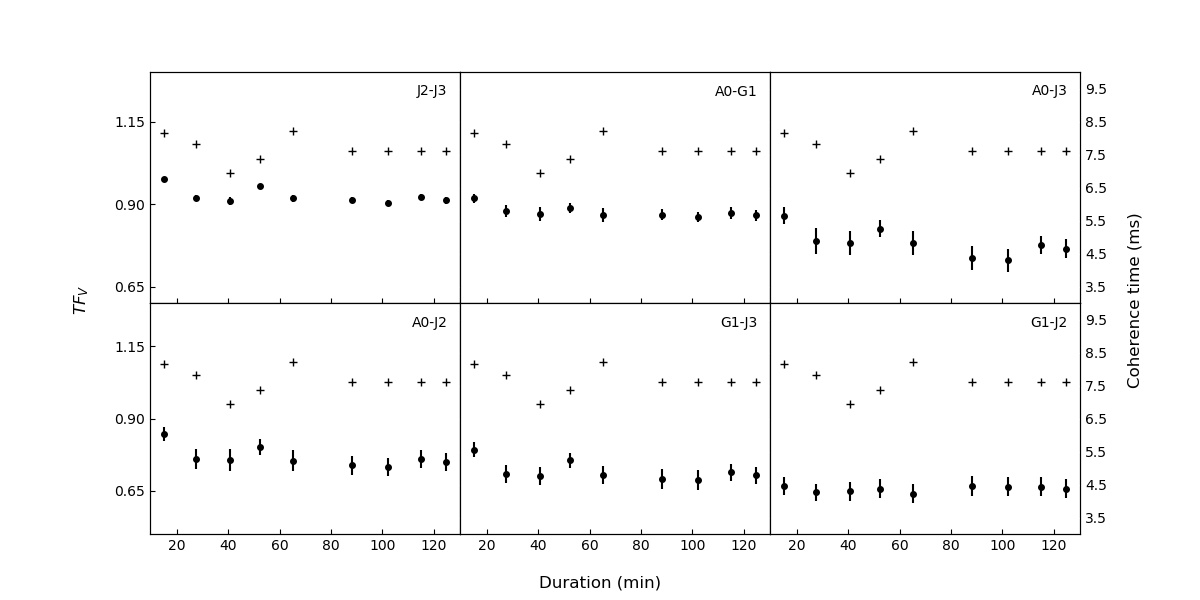}
    \caption{Mean transfer functions in visibility recorded over the night of 2019-07-20 (BCD IN-IN) for the various baselines. The crosses represent the mean coherence time of the atmosphere in the visible.}
    \label{fig:trans_temp}
\end{figure*}

In Fig.~\ref{fig:trans_temp}, we show the mean transfer functions for one BCD state over the 2019 July 20 night, during which nine observations were conducted. The crosses are the mean visible coherence time $\tau_0$ of the atmosphere measured during each data recording. The values show a good and stable situation during the 2 h of observation, i.e. $\tau_0$ of 7.6$\pm$0.4\,ms, and a seeing of 0.56$\pm$0.04\,arcsec. 

The standard deviation of $\rm TF_{V}$ quantifies the stability over the observation. Over the different longest baselines ($\geq$~90\,m) and BCD states, we calculated there is a mean improvement of more than 10 per cent (with a maximum of 14 per cent) if obtained with the present diameters compared to the stability obtained with the JSDC diameters.

Five and six calibrators were consequently observed during 3 other nights. We don't find any significant changes on the stability of the mean function transfer in visibility during these observations, considering our diameter estimations or those reported in the JSDC. We suspect the transfer function stability to be affected by the coherence time variation during these 3 nights, ranging from less than 3\,ms to about 5\,ms in the best conditions. During the MATISSE commissioning (\citet[][]{Petrov2020}), we noticed that the response of the instrument is very sensitive to $\tau_0$ and that a loss in instrumental visibility can exceed a factor 2 when we move from fair conditions ($\tau_0$ > 5\,ms) to bad conditions ($\tau_0$ < 3\,ms).

\section{Calibrator "inspection": 75~Vir analysis} \label{sec:inspection}
Our method can be a good indicator for calibrator reliability. The giant star 75~Vir was initially identified as such (see Table~\ref{tab:calib}). When running our programmes, no stellar diameter results fit with the hypothesis of a UD. We thus investigated more thoroughly the morphology of that star.

\begin{figure*}
\centering
    \includegraphics[width=0.7\textwidth]{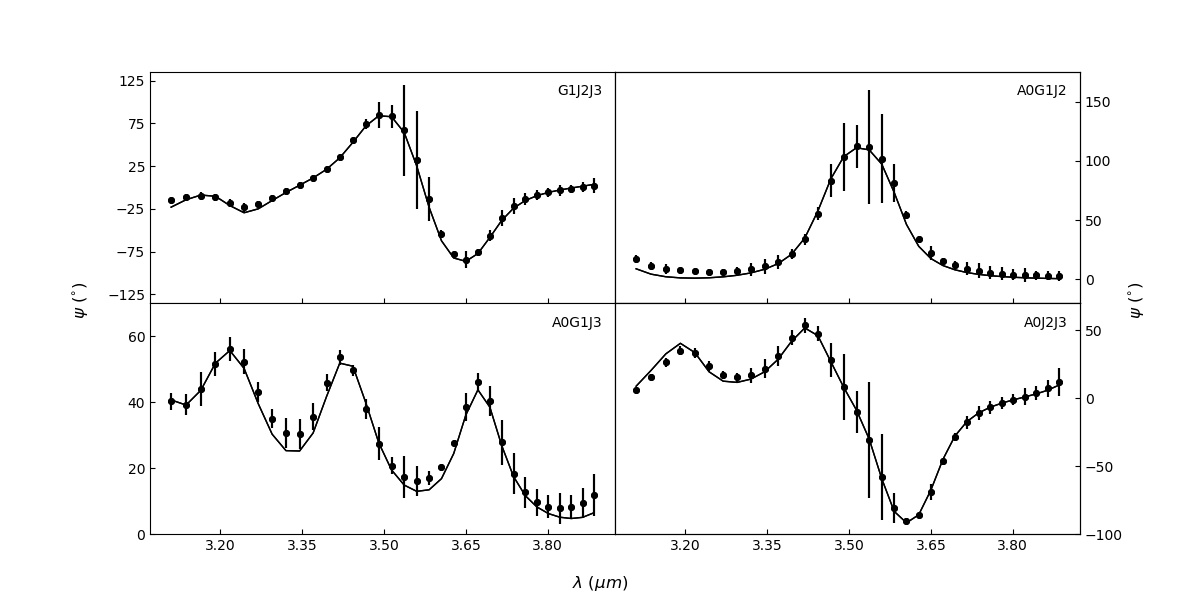}
    \includegraphics[width=0.8\textwidth]{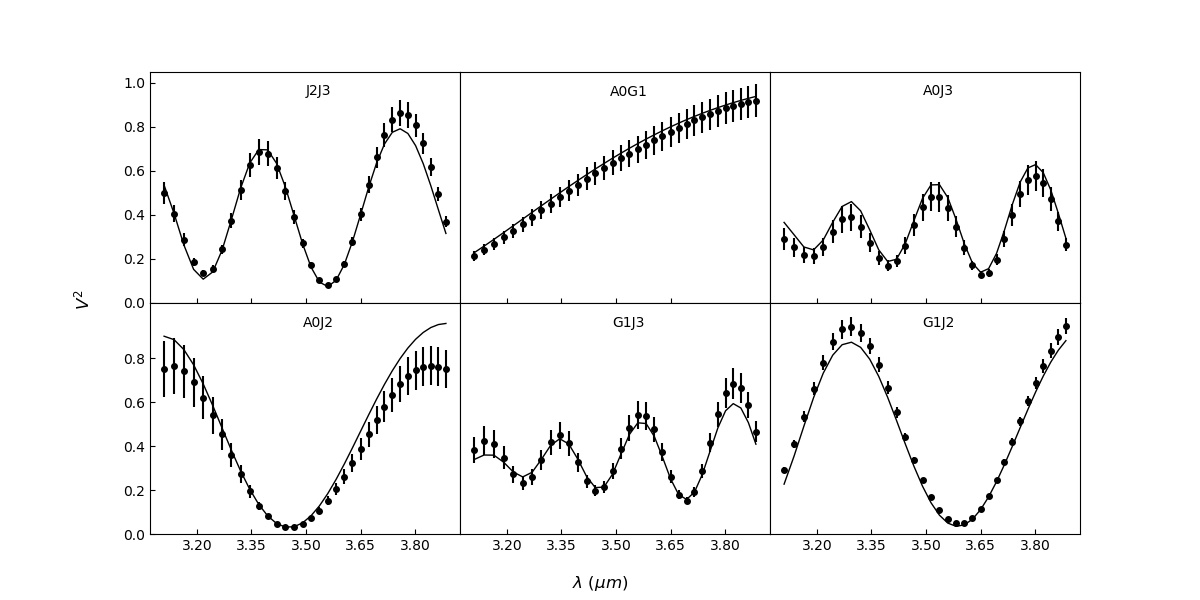}
    \caption{The giant star, 75~Vir, fit of the closure phase $\psi$ and the squared visibility $V^2$ with a binary model.}
    \label{fig:75Vir}
\end{figure*}

Fig~\ref{fig:75Vir} shows the MATISSE $L$ band closure phase $\psi$ (2019 May 23) fitted with a binary model of two point-like stars. The calibration of the squared visibility $V^2$ was done with the giant star 106 Vir. The study ended up with the following relative coordinates $X$ and $Y$ of the two components  \citep[][]{bonneau2015}, resulting in an angular separation $\rho$ and a position angle  $\theta$ [measured from the North (0$^\circ$) to the East (90$^\circ$)]: 
\begin{itemize}
\item $X$~=~33.737$\pm$0.006\,mas;
\item $Y$~=~79.158$\pm$0.007\,mas;
\item $\rho$~=~86.048$\pm$0.009\,mas;
\item $\theta$~=~66.916$\pm$0.005\,$^\circ$. 
\end{itemize}

The same analysis performed with the code CANDID (Companion Analysis and Non-Detection in Interferometric Data) developed by \citet[][]{Gallene2015} ended up with angular separation and position angle very close to our estimations:
\begin{itemize}
\item $X$~=~33.776$\pm$0.035\,mas;
\item $Y$~=~78.967$\pm$0.034\,mas;
\item $\rho$~=~85.887$\pm$0.045\,mas;
\item $\theta$~=~66.84$\pm$0.03\,$^\circ$.
\end{itemize}

\citet[][]{kervella2019} also identified the presence of a companion orbiting 75 Vir. Using the HIPPARCOS and GAIA's second data release (GDR2), a proper motion anomaly was detected, indicative of the presence of a perturbing secondary object. Using the parallax value of 6.58\,mas provided by the GAIA Archives at ESA\footnote{https://gea.esac.esa.int/archive/}, and the value of 86\,mas for the angular separation, we estimate the separation of the stellar components at about 13 AU, providing a possible mass of the companion around 260\,$\rm M_J$, i.e. 0.25~$\rm M_\odot$ \citep[][]{kervella2019}. 


\section{Conclusion}
We propose a new 'self-calibration method' providing stellar diameters of MATISSE calibrators with a precision of 0.6 per cent to 4.1 per cent. Compared to the precision of about 10 per cent provided by the JSDC Stellar Catalogue, we can be more confident in the issued astrophysical results. 

We then select two science targets observed during one commissioning night with good atmospheric conditions and compare the stability of the transfer function in visibility when using our diameter estimations for the calibrators in comparison to that using the JSDC diameter values.
There is an improvement by a factor of two.

We also study the mean transfer function stability in visibility over one stable night of observation of eight calibrators. By comparing the results produced with the JSDC diameters, it is improved with our diameter estimations.

When analysing the selected calibrators, we find that our method does not converge for one of them, 75 Vir, which was previously identified as a binary star. This brings us to the proposition that our method can be used to identify what we can call 'bad calibrators'. Our algorithm will be implemented in the global MATISSE data processing pipeline, available for all users, in order to flag these bad calibrators. That is of high importance for a proper analysis and interpretation of IR interferometric data.

We plan to submit a new ESO proposal to extend this work to bright $N$ band calibrators for ATs, and to bright hybrid calibrators for UTs.

\pagestyle{empty}
\begin{table*}
\renewcommand\thetable{1}
\caption{The list of the stars proposed for our observing program sorted according to the increasing right ascension. The data are reported in the II/361/mdfc-v10 catalogue. The angular diameter $\phi_{\rm JSDC}$ is in mas and the $L$ band flux is in Jy.}            
\label{tab:calib}                              
\begin{tabular}{ ccccccc }        
\hline \hline                
NAME & SPTYPE & RAJ2000 & DEJ2000 & $\phi_{\rm JSDC}$  & IRFLAG &  LFLUX\\
\hline
HD~1089	& K3III	& 00:14:58.3 & -34:54:15 & $1.24\pm0.13$ & 0 &	$20.5\pm1.5$\\
HD~1923	& M2III & 00:23:22.1 & -29:50:50 & $2.39\pm0.19$ & 0 &	$39\pm4$\\
HD~3605	& K4III & 00:38:48.8 & -25:06:28 & $1.32\pm0.13$ & 0 &	$20.9\pm0.2$\\
$\mu$~Psc &	K3III &	01:30:11.1 & +06:08:38 &  $2.46\pm0.24$ & 3	& $75\pm5$\\
$\nu$~Psc & K3III & 01:41:25.9 & +05:29:15 & $2.77\pm0.29$ & 6 & $96\pm4$\\
$\chi$~Eri & G8IIIb & 01:55:57.5 & -51:36:32 & $2.06\pm0.21$ & 0 & $71\pm12$\\
HD~15652 & M0III & 02:30:32.8 &	-22:32:43 & $2.26\pm0.19$ &	0 &	$36\pm8$\\
$\nu$~Cet & G8III & 02:35:52.5 & +05:35:36 & $1.19\pm0.12$ & 0 & $22.2\pm0.3$\\
HD~17973 & M... & 02:53:38.5 & +14:40:13 & $1.91\pm0.16$ & 0 & $23\pm10$\\
HD~19723 & K2 &	03:10:49.2 & +10:00:21 & $1.92\pm0.13$ & 0 & $29\pm6$\\
HD~20356 & K4III & 03:16:21.7 &	-05:43:50 &	$1.88\pm0.18$ &	0 &	$28\pm4$\\
28~Hya	& K4/5III &	09:25:24.0 &	-05:07:03 &	$2.17\pm0.22$ & 0 & $51\pm14$\\
18~Leo &	K4III	& 09:46:23.3 &	+11:48:36 &	$1.96\pm0.19$ &	0	&	$39\pm2$\\
26~Sex	& K3III &	10:26:36.9 &	-00:59:17 &	$1.32\pm0.12$ &	0	&	$22\pm2$\\
HD~90798	& K4III &	10:28:02.1 &	-49:24:21 &	$1.69\pm0.15$ &	0 &	$25\pm9$\\
HD~94014 &	K4/5III &	10:51:05.4 &	-03:05:34 &	$1.60\pm0.15$ &	0	&	$27\pm2$\\
$\iota$~Ant	& K0III	& 10:56:43.1 &	-37:08:16 &	$1.76\pm0.18$ & 0 &	$38\pm5$\\
35~Vir &	M0/1III &	12:47:51.4 &	+03:34:22 &	$2.54\pm0.22$ & 0 & $52\pm23$\\
75~Vir &	K1.5IIIb: &	13:32:51.6 &	-15:21:45	& $1.61\pm0.13$ & 0 & $27.4\pm0.6$\\
106~Vir &	K4/5III &	14:28:41.7 &	-06:54:02	& $2.39\pm0.21$ & 0	&	$55\pm5$\\
$\nu$~Lib &	K5-III	& 15:06:37.6	& -16:15:25	& $2.64\pm0.30$ &	7 &	$78\pm3$\\
HD~135367 &	K3III &	15:14:50.6 &	-05:30:09	& $1.47\pm0.14$ &	0 &	$21.0\pm0.3$\\
37~Lib	& K1III-IV	& 15:34:10.7 &	-10:03:52	& $1.76\pm0.16$ &	0	&	$38\pm4$\\
HD~138938 &	K5III &	15:36:12.9 &	-28:59:56 &	$1.95\pm0.16$ &	0	&	$22\pm6$\\
HD~145085	& K4III	& 16:08:58.9 &	+03:27:16 &	$1.77\pm0.17$ & 0 &	$34.1\pm0.2$\\
HD~153113 &	K0	& 16:57:20.2 &	+06:12:41 &	$1.19\pm0.12$ &	0 &	$19\pm3$\\
V849~Ara	& M3/4Ib &	17:00:55.7 &	-56:28:48 & $2.87\pm0.27$	&	0 &	$40\pm10$\\
HD~162774 &	K4III &	17:52:35.4 &	+01:18:18 & $2.36\pm0.23$ &	0	& $47\pm15$\\
$\eta$~Ser &	K0III-IV	& 18:21:18.6 &	-02:53:56 & $3.10\pm0.33$ & 7 &	$129\pm43$\\
e01~Sgr &	K1III	& 19:40:43.4 &	-16:17:36 & $1.43\pm0.12$ &	0 &	 $28\pm4$\\
HD~184996 &	M0III &	19:41:37.3 &	-65:51:16 &	$2.70\pm0.20$ & 0	& $37\pm4$\\
HD~186251 &	M2III &	19:47:05.8	& -62:27:13 &	$2.25\pm0.18$ & 0	& $31\pm2$\\
HD~189695 &	K5III &	20:00:59.0 &	+08:33:28	& $2.01\pm0.16$ & 0 &	$39\pm8$\\
LU~Aqr &	K3III	& 23:28:07.3	& -12:12:44 &	$2.78\pm0.23$ &	0	&	$38.9\pm1.5$\\
104~Aqr &	G2Ib/II &	23:41:45.8 &	-17:48:60 &	$1.05\pm0.09$ &	0 &	$18.7\pm0.4$\\
\hline
\end{tabular}
\end{table*}
\pagestyle{plain}

\pagestyle{empty}

\begin{table*}
\renewcommand\thetable{2}
\caption{List of the observations carried out from 2019 May to 2020 February with the values of the angular diameter deduced from our method (in mas). The precision $\Delta\phi/\phi$ is the result of the bootstrapping procedure.}            
\label{tab:result}                              
\begin{tabular}{ cccccc  }        
\hline \hline                
DATE & TIME START & BASE CONFIG. & NAME & $\phi$ & $\Delta\phi/\phi$ \\
 &  &  &  & &  (per cent)\\
\hline
2019-05-19 & 04:52:45 & A0--G1--J2--J3 & V849~Ara	&	$2.812\pm0.092$ &	3.3\\
2019-05-19 & 09:37:51 & A0--G1--J2--J3 & e01~Sgr &	$1.430\pm0.028$ &	1.9\\
2019-05-19 & 09:51:08 & A0--G1--J2--J3 & HD~189695 &	$2.041\pm0.059$ &	2.9\\
2019-05-19 & 10:05:22 & A0--G1--J2--J3 & HD~189695 &	$2.023\pm0.069$ &	3.4\\
2019-05-19 & 10:19:21 & A0--G1--J2--J3 & HD~186251 &	$2.226\pm0.037$ & 1.7\\
2019-05-23 & 03:11:19 & A0--G1--J2--J3 & 106~Vir &	$2.381\pm0.030$ &	1.3\\
2019-05-23 & 03:23:38 & A0--G1--J2--J3 & $\nu$~Lib &	$2.599\pm0.035$ &	1.3\\
2019-05-23 & 03:36:08 & A0--G1--J2--J3 & HD~135367 &	$1.459\pm0.019$ &	1.3\\
2019-05-23 & 03:51:39 & A0--G1--J2--J3 & 35~Vir &	$2.510\pm0.053$ &	2.1\\
2019-05-23 & 04:13:54 & A0--G1--J2--J3 & 37~Lib	&	$1.741\pm0.017$ &	0.9\\
2019-05-23 & 04:27:28 & A0--G1--J2--J3 & HD~138938 &	$1.907\pm0.031$ &	1.6\\
2019-07-20 & 08:14:55 & A0--G1--J2--J3 & LU~Aqr & $2.710\pm0.057$ &	2.1\\
2019-07-20 & 08:27:32 & A0--G1--J2--J3 & 104~Aqr &	$1.062\pm0.016$ &	1.5\\
2019-07-20 & 08:40:28 & A0--G1--J2--J3 & HD~1089 & $1.248\pm0.016$ & 1.3\\
2019-07-20 & 08:52:30 & A0--G1--J2--J3 & HD~1923 & $2.366\pm0.035$ & 1.5\\
2019-07-20 & 09:05:03 & A0--G1--J2--J3 & HD~3605 & $1.331\pm0.018$ &	1.4\\
2019-07-20 & 09:28:11 & A0--G1--J2--J3 & $\mu$~Psc & $2.461\pm0.020$ &	0.8\\
2019-07-20 & 09:42:13 & A0--G1--J2--J3 & $\nu$~Psc & $2.772\pm0.021$ & 0.8\\
2019-07-20 & 09:54:58 & A0--G1--J2--J3 & $\chi$~Eri	&$2.067\pm0.015$ &	0.7\\ 
2019-07-20 & 10:04:50 & A0--G1--J2--J3 & $\chi$~Eri	& $2.055\pm0.016$ &	0.8\\
2019-07-21 & 07:29:37 & A0--G1--J2--J3 & HD~184996 &	$2.656\pm0.024$ &	0.9\\
2019-07-21 & 09:20:13 & A0--G1--J2--J3 & $\mu$~Psc & $2.462\pm0.022$ &	0.9\\
2019-07-21 & 09:32:46 & A0--G1--J2--J3 & HD~15652 & $2.225\pm0.037$ &	1.7\\
2019-07-21 & 01:15:13 & A0--G1--J2--J3 & HD~145085	&	$1.782\pm0.038$ &	2.1\\
2019-07-21 & 01:27:58 & A0--G1--J2--J3 & HD~153113 &	$1.224\pm0.021$ &	1.7\\
2019-07-21 & 01:54:11 & A0--G1--J2--J3 & HD~162774 &	$2.390\pm0.095$ &	4.0\\
2019-07-23 & 07:07:19 & A0--G1--J2--J3 & HD~184996 &	$2.676\pm0.016$ &	0.6\\
2019-08-21 & 09:15:40 & A0--G1--J2--K0 & HD~19723	& $1.930\pm0.015$  &	0.8\\
2019-08-24 & 09:00:11 & A0--G1--J2--J3 & $\nu$~Cet & $1.198\pm0.016$ &	1.3\\
2019-08-24 & 09:15:49 & A0--G1--J2--J3 & HD~17973 &	$1.941\pm0.019$ &	1.0\\
2019-08-24 & 09:29:25 & A0--G1--J2--J3 & HD~20356 &	$1.901\pm0.029$ &	1.5\\
2019-08-25 & 03:28:08 & A0--G1--J2--J3 & $\eta$~Ser &	$3.078\pm0.128$ &	4.1\\
2019-12-31 & 05:49:51 & A0--G1--J2--K0 & HD~90798 &	$1.709\pm0.020$ & 1.2\\
2020-02-02 & 08:33:31 & A0--B2--D0--J3 & HD~94014 &	$1.592\pm0.025$	& 1.6\\
2020-03-01 & 03:46:27 & A0--G1--J2--J3 & 28~Hya	&	$2.151\pm0.031$ &1.4\\
2020-03-01 & 04:01:27 & A0--G1--J2--J3 & 26~Sex	&	$1.314\pm0.014$ &	1.1\\
2020-03-01 & 05:04:24 & A0--G1--J2--J3 & 18~Leo & $1.962\pm0.020$ &	1.0\\
2020-03-12 & 05:04:24 & U1--U2--U3--U4 & $\iota$~Ant & $1.754\pm0.016$ & 0.9\\
\hline
\end{tabular}
\end{table*}
\pagestyle{plain}

\pagestyle{empty}
\begin{table*}
\renewcommand\thetable{3}
\caption{Comparison of the results between the linearly LD model for values of the limb darkening parameter $u$ from 0 to 1, and the UD model. The precision $\Delta\phi_{\rm UD}/\phi_{\rm UD}$ of the method obtained for the UD (Table~\ref{tab:result}), with $\phi_{\rm UD}~=~\phi$ the UD angular diameter, is referred. $\phi_0$ and $\phi_1$ are respectively the fitted LD angular diameters at $u~=~0$ and at $u~=~1$. $\sigma$ is the ratio between the root mean square error of the angular diameter distribution and $\phi_0$.}            
\label{tab:ass}                              
\begin{tabular}{ ccccccc  }        
\hline \hline                
NAME & $\Delta\phi_{\rm UD}/\phi_{\rm UD}$& $\phi_0$&  $\phi_1$  &$(\phi_0-\phi_{\rm UD})/\phi_{\rm UD}$ & $\sigma$ & $(\phi_1-\phi_0)/\phi_0$ \\
&(per cent)& (mas) & (mas)& (per cent)& (per cent)& (per cent)\\
\hline
V849~Ara & 3.3& 2.8061 &2.8097 &	-0.2& 0.6 &	 0.1  \\
e01~Sgr & 1.9& 1.4345& 1.4351&0.3&	0.3&	0.04\\
HD~189695& 2.9& 2.0341& 2.0278&-0.3&	0.3&	-0.3\\
HD~186251& 1.7& 2.2290&2.2470 &	0.1&	0.3&	0.8\\
106~Vir& 1.3& 2.3767 & 2.3876&-0.2&	0.5&	0.5\\
$\nu$~Lib&1.3 & 2.5998& 2.6353 &0.02&	0.2&	1.4\\
HD~135367& 1.3&1.4622 & 1.4731&	0.2&	0.3&	0.8\\
35~Vir &2.1& 2.5114& 2.5362	&0.1 &0.2&	1.0\\
37~Lib&	0.9& 1.7380&1.7468 &-0.2&	0.2&	0.5\\
HD~138938 &1.6 &1.9103 & 1.9207&	0.2&	0.3	&0.5\\
LU~Aqr & 2.1& 2.7120& 2.7578& 0.1&	0.2&	1.7\\
104~Aqr & 1.5&1.0625 &1.0676 &	0.1&	0.4&	0.5\\
HD~1089 &1.3 &1.2475 &1.2451 &-0.02&	0.3	&-0.2\\
HD~1923 & 1.5&2.3667 & 2.3917&0.01&	0.2&	1.1\\
HD~3605 &1.4 & 1.3344&1.3284 &-0.2&	0.3& -0.5\\
$\mu$~Psc& 0.8&2.4608 &2.4698 &-0,1&	0.1	&0.4\\
$\nu$~Psc& 0.8&2.7846 &2.7953 & 0.5&	0.2	&0.4\\
$\chi$~Eri&	0.7& 2.0606& 2.0690& 0.3&	0.2&	0.4\\
HD~184996& 0.6&2.6691&	2.6886&  -0.2&	0.1	&0.7\\
HD~15652&1.7 &2.2270 &2.2558 & 0.1	&0.3&	1.3\\
HD~145085&2.1&1.7762 &1.7843	&	-0.3&	0.2&	0.5\\
HD~153113&1.7 &1.2192 &1.2184 &-0.4	&0.5&	-0.04\\
HD~162774& 4.0& 2.3856&2.3880 &	-0.2&	0.2&	0.1\\
HD~19723	&0.8 &1.9264 &1.9305 & -0.2&	0.2&	0.2\\
$\nu$~Cet& 1.3&1.1945 & 1.1944& -0.2&	0.4&	-0.04\\
HD~17973 & 1.0& 1.9324 &1.9391 &-0.4&	0.2&	0.4\\
HD~20356 & 1.5&	1.8975&1.9054 &-0.2&	0.2&	0.4\\
$\eta$~Ser&4.1 &3.0842 &3.1267 &0.2	&0.2&	1.4\\
HD~90798& 1.2&1.6980	&1.7091 &-0.6&	0.2&	0.7\\
HD~94014 &1.6&1.5902	&1.6231 &-0.1&	0.9&	2.1\\
28~Hya	&1.4& 2.1534&2.1625 & 0.1&	0.2&	0.4\\
26~Sex	&1.1&1.3128 &1.3238 &-0.1&	0.3&	0.8\\
18~Leo &1.0 &1.9559 & 1.9666&-0.3&	0.3&	0.5\\
$\iota$~Ant& 0.9& 1.7467& 1.7518 & -0.4&	0.3&	0.3\\
\hline
\end{tabular}
\end{table*}
\section*{Acknowledgements}

MATISSE has been built by a consortium composed of
French (INSU-CNRS in Paris and OCA in Nice), German (MPIA, MPIfR and
University of Kiel), Dutch (NOVA and University of Leiden), and Austrian (University
of Vienna) institutes. It was defined, funded and built in close collaboration
with ESO. The Conseil D\'epartemental des Alpes-Maritimes in France, the
Konkoly Observatory and Cologne University have also provided resources for
the manufacture of the instrument. We have great remembrance of the contributions
from our two deceased colleagues, Olivier Chesneau and Michel Dugu\'e.
This work was supported by the Action Sp\'ecifique Haute R\'esolution Angulaire
(ASHRA) of CNRS/INSU co-funded by CNES, as well as from the Programme
National de Physique Stellaire (PNPS) from CNRS.

This work is based on observations collected at the European Southern Observatory, Chile (ESO ID: 103.L-0941).

\section*{Data availability}
The raw data underlying this work are publicly available at http://archive.eso.org using the ESO program ID: 103.L-094. The processed observational data are available at http://oidb.jmmc.fr/index.html, by selecting the MATISSE OIFITS files, in the collection 'LM-band interferometric calibrators - L1 Level'.









\bsp	
\label{lastpage}
\end{document}